\documentclass[sigconf,natbib=true]{acmart}

\usepackage{hyperref}
\usepackage{url}
\usepackage{wrapfig}
\usepackage{graphicx}
\usepackage{subcaption} 
\usepackage{multirow}
\usepackage{makecell}
\usepackage[table]{xcolor}
\usepackage{tabularx}
\definecolor{mygold}{RGB}{230,190,70}
\usepackage{balance}
\usepackage{minted}
\definecolor{LightGray}{gray}{0.9}
\setminted{
    baselinestretch=0.8,
    bgcolor=LightGray,
    fontsize=\footnotesize,
}
\setminted[python]{breaklines}
\usepackage{enumitem}

\usepackage[dvipsnames]{xcolor}
\usepackage{pifont}   
\newcommand{\cmark}{\textcolor{ForestGreen}{\ding{51}}}

\usepackage{booktabs}
\usepackage{threeparttable}
\usepackage{siunitx}

\newcommand{\NA}{\textemdash}

\AtBeginDocument{%
  }

\begin{document}
\title[SplitLight: An Exploratory Toolkit for Recommender Systems Datasets and Splits]{SplitLight: An Exploratory Toolkit \\ for Recommender Systems Datasets and Splits}

\author{Anna Volodkevich}
\orcid{0009-0002-7958-0097}
\affiliation{%
  \institution{SB AI Lab, AI Center}
  \country{}
}

\author{Dmitry Anikin}
\orcid{0009-0001-9715-0616}
\affiliation{%
  \institution{SB AI Lab, Applied AI Institute}
  \country{}
}

\author{Danil Gusak}
\orcid{0009-0008-1238-6533}
\affiliation{%
  \institution{AXXX, AI Center}
  \country{}
}

\author{Anton Klenitskiy}
\orcid{0009-0005-8961-6921}
\affiliation{%
  \institution{SB AI Lab}
  \country{}
}

\author{Evgeny Frolov}
\orcid{0000-0003-3679-5311}
\affiliation{%
  \institution{AXXX, HSE University}
  \country{}
}

\author{Alexey Vasilev}
\orcid{0009-0007-1415-2004}
\affiliation{%
  \institution{SB AI Lab, MSU Research Center}
  \country{}
}

\acmDOI{XXXXXXX.XXXXXXX}
\copyrightyear{2026}
\acmYear{2026}
\setcopyright{none}

\acmConference[Conference acronym '26]{X}{2026}{XXX | X, X}
\acmBooktitle{X, 2026, XXX | X, X}



\begin{teaserfigure}
\centering
\includegraphics[width=0.95\textwidth]{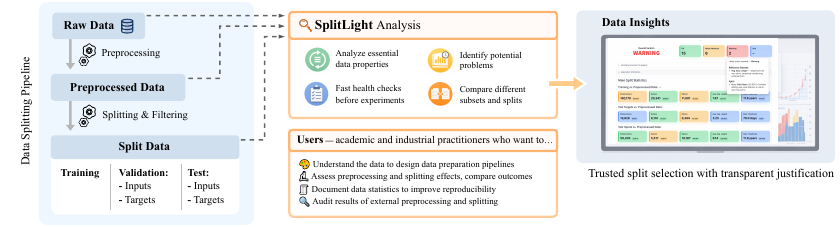}
\caption{\textbf{SplitLight} in a data-preparation pipeline. From the raw dataset to split subsets, SplitLight audits data characteristics, flags problems, and enables side-by-side comparison of alternative splitting strategies to justify the selected evaluation protocol.}
\label{fig1}
\end{teaserfigure}
\begin{abstract}

Offline evaluation of recommender systems is often affected by hidden, under-documented choices in data preparation. Seemingly minor decisions in filtering, handling repeats, cold-start treatment, and splitting strategy design can substantially \textit{reorder model rankings} and \textit{undermine reproducibility and cross-paper comparability}.

In this paper, we introduce \textit{SplitLight}, an open-source exploratory toolkit that enables researchers and practitioners designing preprocessing and splitting pipelines or reviewing external artifacts to make these decisions \textit{measurable, comparable, and reportable}.
Given an interaction log and derived split subsets, SplitLight analyzes core and temporal dataset statistics, characterizes repeat consumption patterns and timestamp anomalies, and diagnoses split validity, including temporal leakage, cold-user/item exposure, and distribution shifts.
SplitLight further allows side-by-side comparison of alternative splitting strategies through comprehensive aggregated summaries and interactive visualizations.
Delivered as both a Python toolkit and an interactive no-code interface, SplitLight produces audit summaries that justify evaluation protocols and \textit{support transparent, reliable, and comparable} experimentation in recommender systems research and industry.

\begin{center}
  SplitLight: \href{https://github.com/monkey0head/SplitLight}{%
    \textcolor{purple}{https://github.com/monkey0head/SplitLight}%
}
\end{center}
\end{abstract}

\maketitle

\section{Introduction}

Reliable offline evaluation of recommender systems remains a high-impact yet under-controlled problem in both research and production practice. Every evaluation pipeline starts with \textit{data preprocessing and splitting}, and prior work shows that these preliminary steps have a \textit{strong influence on the evaluation results}~\cite{gusak2025time,hidasi2023flaws}. Decisions that often look minor  -- filtering criteria, treatment of repeats and cold items, or what exactly is provided as test input and held out as target -- can substantially reorder model rankings and unfairly favor certain algorithms~\cite{klenitskiy2026analysis,meng2020splits,cold2024}. When these design choices are opaque or inconsistently reported, results become hard to reproduce and nearly impossible to compare across studies.

A principled evaluation, therefore, requires \textit{auditing the data both before and after preprocessing and validating the post-split subsets} that are ultimately used for training and evaluation. At the dataset level, this includes not only standard core statistics but also temporal characteristics and repeat consumption patterns. At the split level, data splitting results should be checked for data sufficiency and realism, temporal overlaps and leakage~\cite{gusak2025time,hidasi2023flaws,ji2020leakage,le2025don}, cold-start exposure~\cite{gusak2025dish, pembek2025let, cold2024, sukhorukov2025maximum}, and distributional shifts between training and evaluation data~\cite{sun2023take}. Being useful for universal recommender pipelines, such analysis is especially acute for sequential recommendation~\cite{volodkevich2025autoregressive,mezentsev2024scalable,khrylchenko2025scaling}, where timestamp errors~\cite{fan2024our}, sessionization choices, or temporal leakage significantly distort user state and inflate next-item metrics~\cite{hidasi2023flaws,le2025don,ludewig2018sessioneval}.
.

Existing work on recommendation datasets emphasizes data management and curation pipelines, benchmarking variability, and high-level dataset property analysis~\cite{klenitskiy2024does,mancino2025datarec,shevchenko2024variability}. However, it does not focus on data preprocessing and splitting steps and may only cover individual aspects such as reporting basic dataset characteristics~\cite{mancino2025datarec}. While researchers analyzing datasets or investigating causes of unexpected experimental results, identify and highlight important data properties and quality issues, including timestamp collisions \cite{hidasi2023flaws}, user lifetime distribution properties \cite{fan2024our, klenitskiy2026analysis}, post-split data sufficiency and evaluation subset skew \cite{gusak2025time, sun2023take, ji2020leakage, meng2020splits}, \textit{there is still no comprehensive, actionable "what to inspect" list and no unified instrument for performing such analysis}.

In practice, researchers and practitioners repeatedly face the need to: (i) understand the data to design preprocessing and splitting pipelines; (ii)  analyze, how different preprocessing and splitting steps affect the data, and compare the outcomes; (iii)  document statistics to enhance reproducibility; and (iv) audit the results of external (black-box) preprocessing and splitting pipelines to identify and fix potential issues. In this work, we address these needs with a comprehensive auditing checklist and an easy-to-use tool.

We distill practical lessons from building recommender systems pipelines together with relevant prior research, highlighting particular elements of data curation setups and their potential impact on evaluation outcomes. We identify and summarize the essential properties of datasets and data splits, and we make the results available to the community with \textbf{SplitLight}, a lightweight open-source exploratory toolkit that provides indicators and tools to \textit{diagnose key data properties and assess preprocessing and splitting results}.
While SplitLight is designed around common recommender systems interaction logs, the same auditing principles and interfaces apply to other domains with user and item identifiers and timestamps, including information retrieval (IR) logs and transaction data.

In summary, the main contributions of our work are:
\begin{itemize}[leftmargin=*]
    \item We identify and summarize important dataset and split properties that should be analyzed when designing a preprocessing and splitting pipeline;
    \item We introduce \textbf{SplitLight}\footnote{Code is available at: \url{https://github.com/monkey0head/SplitLight}. \href{https://drive.google.com/file/d/15ZSKai7dYXBVmcqPIuV4qsM9bbsRxNY4/view}{\textcolor{purple}{Video walkthrough}.}}, an open-source Python toolkit for auditing datasets and post-split subsets to avoid common issues and enable transparent, reproducible evaluation (Figure~\ref{fig1}). SplitLight provides a no-code interactive Streamlit\footnote{Live demo of the Streamlit interface is available at: \url{https://splitlight.streamlit.app}} interface for rapid analysis, lowering the barrier to rigorous evaluation practice.
    \item We showcase SplitLight on six widely used datasets
to illustrate (i) how the toolkit can audit datasets and data splits, and (ii) how different treatments of identified data properties affect the RS model performance. 

\end{itemize}

\section{Common Dataset and Data Split Properties}

In this section, we outline key attributes of RS and IR datasets we consider and clarify which attributes remain out of scope. We then formalize the split structure and highlight common sources of ambiguity in split specifications. Finally, we describe the essential dataset and data split characteristics, and explain why they matter for reliable evaluation.

\subsection{Key Attributes of RS and IR Datasets}
A user-item interaction log is the core of most RS and IR datasets.
For modern sequential recommender systems in particular, the standard representation is the interaction triple \textit{(user, item, timestamp)}: \textit{who} acted, \textit{what} they acted on, and \textit{when} the event occurred. These three attributes are sufficient to construct ordered user histories and define prediction points for future interactions.
In this work, we deliberately do not require additional features that may improve modeling quality, such as interaction-specific attributes (e.g., amount, rating value, interaction type), contextual signals, or user and item side information, since these are inconsistently available across public datasets and often proprietary in industrial settings.
We consider timestamps essential, as they enable proper ordering of user sequences for sequential recommendation models and help prevent information leakage from the “future” during data splitting.

\begin{figure}
\includegraphics[width=0.45\textwidth]{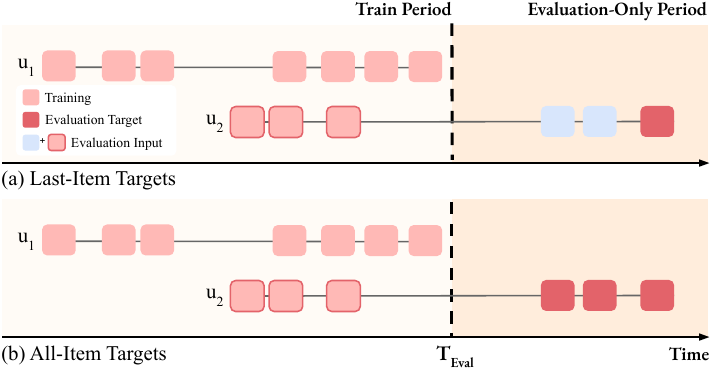}
\caption{Global temporal split structure for (a) last-item and (b) all-items target definitions. Evaluation inputs include all within-user history before evaluation targets.} \label{split}
\end{figure}

\subsection{Data Splitting Protocol Specification}
\label{split_structure} 
After data preprocessing, interaction logs are typically split into training, validation, and test sets for offline model evaluation, and the split type and its main parameters (e.g., leave-one-out split or a global temporal split with an 80/10/10 ratio) are reported. However, this is often not sufficient for reproducibility. The community is pushing toward evaluation pipelines that are both more realistic and more reproducible~\cite{gusak2025time}, which requires reporting more detailed split descriptions. In real use cases, a trained model is applied to future data, including users and items that were not available during training. 
Therefore, it is important to specify the split pipeline more clearly: which data is (1) used for training, (2) used as model input at inference, (3) held out as evaluation target(s), and (4) excluded.  

Thus, we propose to structure splitting results as presented in Figure \ref{split}, illustrated on a global temporal split with last-item and all-items target definitions~\cite{gusak2025time}. We first define the \textit{training} subset. In practice, custom steps such as removing short sequences (e.g., length $1$) or training data augmentation/undersampling \cite{tan2016improved,xie2022cl4srec,sachdeva2022samplingcf} are frequently applied, altering the original interaction logs and, thus, the sequences used for training may differ from the original data. For evaluation, the \textit{test and validation subsets} are further divided into \textit{input and target} parts. The \textit{input} subset contains user history provided to the model at inference time, while the \textit{target} subset includes the ground-truth items the model should predict.

\subsection{Dataset and Split Properties to Audit}

Practitioners and researchers working with recommender systems often face practical questions related to data quality, dataset suitability, and the validity of data splits used for evaluation. These questions arise when assessing whether a dataset is appropriate for a specific task and whether experimental results can be trusted. Common concerns include the following:

\begin{enumerate}[leftmargin=*]
    \item What are the main characteristics of the dataset? Does it contain clear issues or important properties that should be considered during preprocessing and data splitting to obtain reliable training results? Is it similar to the production data and suitable for the intended task? 

    \item What are the properties of the data splits? Are they reasonable in terms of the number of users, interaction volume, and time coverage in the training and evaluation subsets? Is there data leakage from the future?
    
    \item Does the dataset exhibit repeat consumption behavior? How common are repeated interactions, and should they be specifically handled?

    \item Does the chosen splitting strategy introduce a cold-start problem? How severe is this effect in the evaluation subsets, and does it reflect the model's deployment scenario?

    \item Is externally preprocessed and split data reliable and free from major errors?

    \item Are the evaluation subsets representative of the training data, or are they affected by significant bias or skew?

\end{enumerate}

These questions highlight the need for \textit{practical tools} that support \textit{systematic analysis} of recommender system datasets and data splits. Based on these needs, we identify several \textit{key areas of interest} for the dataset and split analysis:
    (i) core and temporal properties of datasets and splits;
    (ii) distribution of interactions over time;
    (iii) repeat consumption patterns;
    (iv) indicators of data leakage;
    (v) cold-start characteristics;
    and (vi) distribution shifts between training and evaluation subsets.
Although this list is not exhaustive, it covers several \textit{essential aspects of data quality and evaluation reliability}. It also provides a foundation that can be extended in future research.

\section{SplitLight}

\textit{SplitLight} is a tool for analyzing datasets and data splitting results. It reveals key data insights, identifies potential issues, and enables side-by-side comparison of alternative splitting strategies to support principled, defensible choices. Such analysis reveals critical properties and problems that must be quantified, mitigated, and explicitly reported in the evaluation protocol to ensure transparent and reproducible RS evaluation. 

\subsection{Availability and Utility}
SplitLight is intended for researchers and practitioners who design preprocessing and splitting pipelines or need to audit externally produced artifacts to ensure they meet specific requirements. 
The toolkit can be used in Jupyter/Python scripts for comprehensive analysis and via a Streamlit UI for rapid, no-code exploration and comparison. SplitLight is released as open-source software under the MIT License and is publicly available on GitHub. The repository provides launch instructions\footnote{\url{https://github.com/monkey0head/SplitLight/blob/main/README.md}}, executable usage examples\footnote{\url{https://github.com/monkey0head/SplitLight/blob/main/demo_Diginetica.ipynb},  \url{https://github.com/monkey0head/SplitLight/blob/main/demo.ipynb}}, and a short \emph{\href{https://drive.google.com/file/d/15ZSKai7dYXBVmcqPIuV4qsM9bbsRxNY4/view}{video walkthrough}}. Streamlit application embeds documentation describing each reported statistic, diagnostic, and the overall framework usage. The codebase is designed to be practical to adapt: users can modify and extend SplitLight to meet specific project needs. The Streamlit UI can be deployed locally from the project directory via \texttt{streamlit run SplitLight.py} command.

\subsection{Data Structure}
SplitLight analyzes and compares data subsets that typically arise in experimental pipelines: \textit{raw} data, \textit{preprocessed} data, and post-split \textit{training} and evaluation data. Evaluation consists of \textit{input} and \textit{target} subsets  (Section \ref{split_structure}) for \textit{validation} and \textit{test}. Data can be provided directly to SplitLight or constructed using its built-in utilities.

\begin{figure}
\setlength{\abovecaptionskip}{6pt}
\includegraphics[width=1.0\columnwidth]{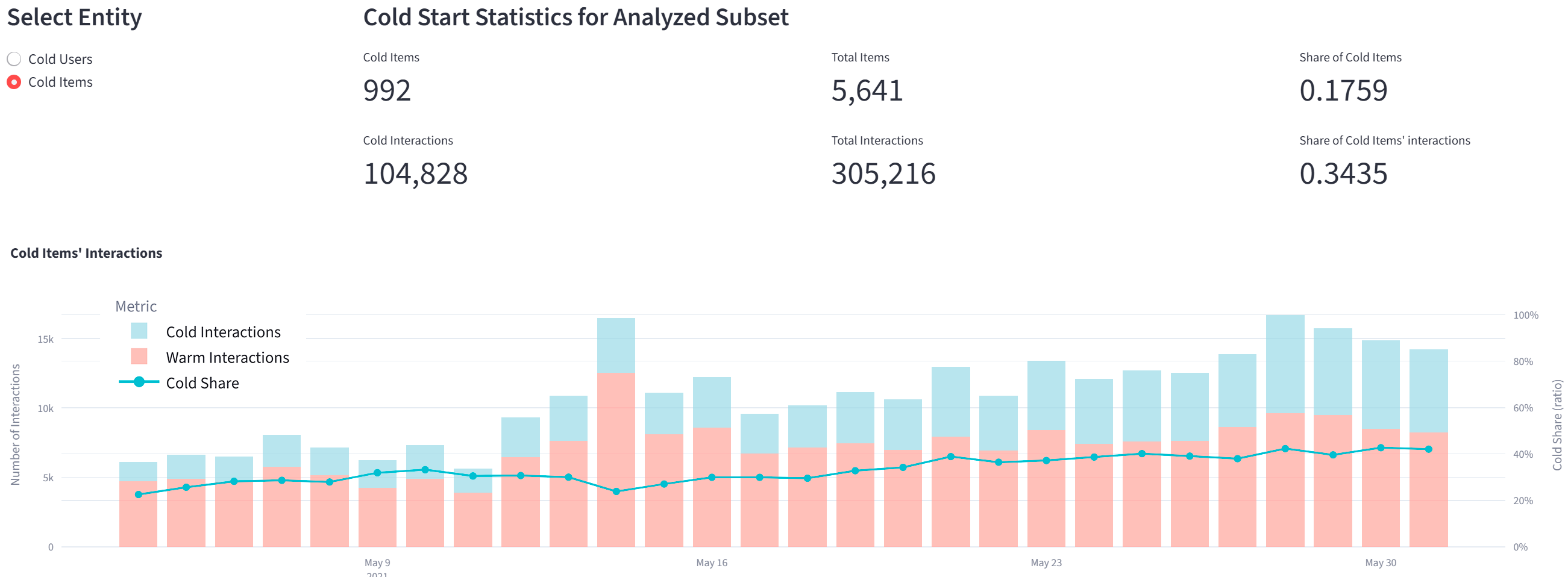}
\caption{Analysis of cold items in test target subset for Diginetica dataset. Share of interactions with cold items grows in time from nearly 20\% to 40\%, resulting in 34.35\% overall.} \label{fig:cold_start}
\end{figure}

\subsection{Dataset-Level Properties}
\subsubsection{Core and Temporal Statistics}\label{sec:core_temporal}

Core dataset properties that are often reported for reproducibility include \textit{number of interactions, users, and items, item popularity, sequence length, and density}. In this work, we highlight another key dimension: \textit{time}. Temporal analysis is needed to properly order user histories, design realistic splits, identify problems in data collection, and spot unexpected temporal shifts. 
Temporal characteristics also help practitioners judge how similar an open dataset is to their in-house data.

\begin{figure}[t]
\setlength{\abovecaptionskip}{7pt}
\includegraphics[width=1.0\columnwidth]{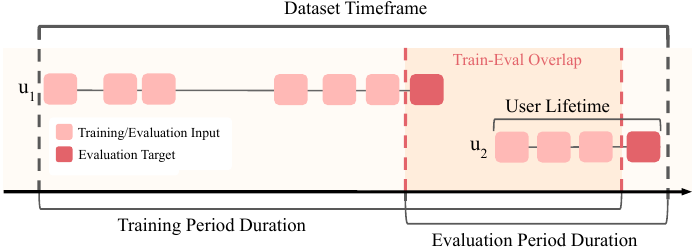}
\caption{Temporal properties of the data in case of the leave-one-out split: dataset timeframe, train/evaluation durations and overlap, and per-user lifetime coverage.} \label{fig:temporal_props}
\end{figure}

\emph{Implementation.} SplitLight offers analysis of core characteristics, such as \textit{number of interactions, users, and items, item popularity, sequence length, and density}, including distributional analysis for per-item/user characteristics. Temporal characteristics include:
\begin{itemize}[leftmargin=*]
    \item \textit{Dataset timeframe}: time between the first and last interaction in the dataset (Figure~\ref{fig:temporal_props}).
    \item \textit{Time delta between interactions}: time between two consecutive interactions of the same user.
    \item \textit{Timestamp collision rate}: how frequently the same timestamp appears within a single user’s history. The presence of timestamp collisions indicates data quality issues and makes time-based splitting non-deterministic.
    \item \textit{User and item lifetime}: time between the user’s (or item’s) first and last interaction (Figure~\ref{fig:temporal_props}). This helps to analyze how quickly the user base and item catalog change over time. 
\end{itemize}
Statistics are computed for the selected subset and, optionally, compared with a reference subset (typically raw or preprocessed data).

\begin{figure}[t]
\setlength{\abovecaptionskip}{7pt}
\includegraphics[width=1.0\columnwidth]{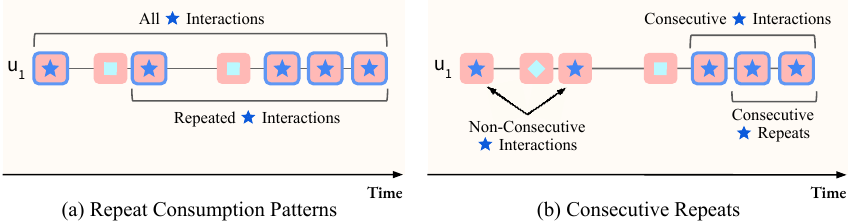}
\caption{Repeat consumption illustration: (a) repeated interactions (after first occurrence of an item) and (b) consecutive repeats (same item in immediate succession).} \label{fig:repeats}
\end{figure}

\subsubsection{Interactions Distribution Over Time.}\label{sec:over_time}

Visual analysis of data distribution over time helps identify potential data-collection issues (e.g., missing data), data skew, and activity fluctuations, and can detect temporal overlaps between split subsets.

\begin{figure}[t]
    \centering
    \resizebox{1.0\columnwidth}{!}{%
    \includegraphics[width=1.1\linewidth]{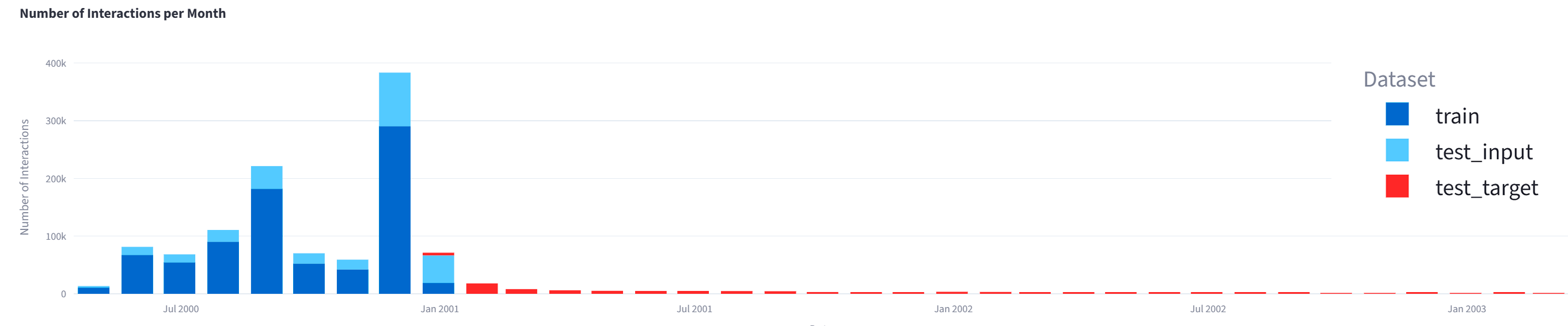}
    }
    \caption{Temporal skew in ML-1M interactions. Test target subset spans nearly 76\% of dataset timeframe under GTS $q_{0.9}$.}
    \label{fig:over_time}
\end{figure}

\emph{Implementation.}
The \textit{Interactions Over Time} functionality visualizes how interactions are distributed over time across different subsets. It aggregates interaction counts at a selectable temporal granularity within a chosen date range, as illustrated in Figure~\ref{fig:over_time}.

\subsubsection{Repeat Consumption Patterns}

Repeated user-item interactions can affect both training behavior and the interpretation of offline evaluation results. Their share depends on the domain: repeats are rare in some datasets (e.g., one-time consumption) but common in streaming and replenishment scenarios~\cite{repeatconsumption2020doi}. 
Consecutive repeats should be analyzed separately, as they can indicate data logging issues or may need to be collapsed or aggregated to improve RS performance (Section \ref{sec:res_model}).

\emph{Implementation.}
SplitLight measures repeat consumption at the user level and reports both absolute counts and relative shares, together with distributions across users. It distinguishes (i) \textit{repeated interactions}, defined as any occurrence of an item after its first appearance in a user’s history, and (ii) \textit{consecutive repeats}, where the same item appears in immediate succession (Figure~\ref{fig:repeats}).

\subsection{Split-Level Properties}
SplitLight evaluates whether a split matches the intended deployment scenario. The analysis focuses on four common risks: \textit{data insufficiency and unrealisticness}, \textit{temporal leakage}, \textit{cold-start exposure}, and \textit{distribution shift} between training and evaluation subsets.

\subsubsection{Core and Temporal Split Statistics}
Split subsets should be reasonable in terms of the number of users, interaction volume, and time coverage in the training and evaluation subsets. More realistic, but more complex splitting strategies, such as global temporal split with target item selection, require dataset-specific quantile selection to match real-world usage scenarios and assure data sufficiency~\cite{gusak2025time}. A side-by-side comparison helps justify the selected protocol and improves reproducibility.

\emph{Implementation.}
SplitLight functionality, described in Sections~\ref{sec:core_temporal}, \ref{sec:over_time}, is also applicable for splitting results analysis. Select evaluation subsets as the \textit{Analysed Subset} and compare with the training or preprocessed data to analyze the main properties of the splitting results. Use the "Compare Splits: Core and Temporal Statistics" functionality to compare splitting protocols and select the best one.

\subsubsection{Temporal Leakage}

Temporal leakage occurs when training data contains information that would not be available at prediction time in real-world settings, since training and evaluation correspond to non-overlapping time periods. This can inflate offline metrics and bias comparisons by favoring methods that can exploit leaked signals~\cite{ji2020leakage, hidasi2023flaws}. Even without direct leakage, when evaluation subset interactions occur earlier than training interactions with the same item, future data can provide insights into trending content types and user preference drift. The goal of the analysis is to ensure the absence of leakage or, when present, to quantify it and assess its potential impact on reported results.

\emph{Implementation.}
SplitLight checks (i) shared interactions between training and evaluation subsets, (ii) temporal overlap between their time ranges (Figure~\ref{fig:temporal_props}), and (iii) \textit{distribution and share of evaluation targets affected by data leakage} (Figure~\ref{fig:leakage}).

\subsubsection{Cold Start}

Cold-start evaluation (users or items not observed during training) measures a different generalization regime than warm-start recommendation~\cite{cold2024, better2024}. Different splitting strategies induce different cold-start rates. If these rates are not controlled, the evaluation protocol may deviate from the intended usage scenario or disproportionately reflect performance on the dominant interaction type (warm or cold). For example, a primarily warm-start evaluation may unintentionally favor methods that rely on learned embeddings over content-based and hybrid approaches.

\emph{Implementation.} SplitLight identifies \textit{cold users} and \textit{cold items} in the validation/test subsets as those not present in the training set. It reports their counts and shares, and tracks how the prevalence of cold interactions changes over time.

\subsubsection{Distribution Shift between Training and Evaluation}

A split is most informative when the evaluation data matches what the model will face after deployment. Split design can introduce systematic distributional shifts across evaluation targets, including changes in \textit{time-gaps} (the time between the last input interaction and the target) and biases in \textit{target position} within sequences. Such split-induced shifts can reduce the alignment between offline and online performance due to differences in distributional properties.

\emph{Implementation.}
SplitLight compares target \textit{time-gap} distributions against a reference distribution (typically all consecutive gaps in the raw or preprocessed data) and quantifies differences using the \textit{Kolmogorov-Smirnov} statistic~\cite{massey1951ks}. It similarly compares \textit{target positions} within user sequences. These outputs help detect and quantify split-induced shifts affecting offline conclusions.

\begin{figure}[t]

\includegraphics[width=1.0\columnwidth]{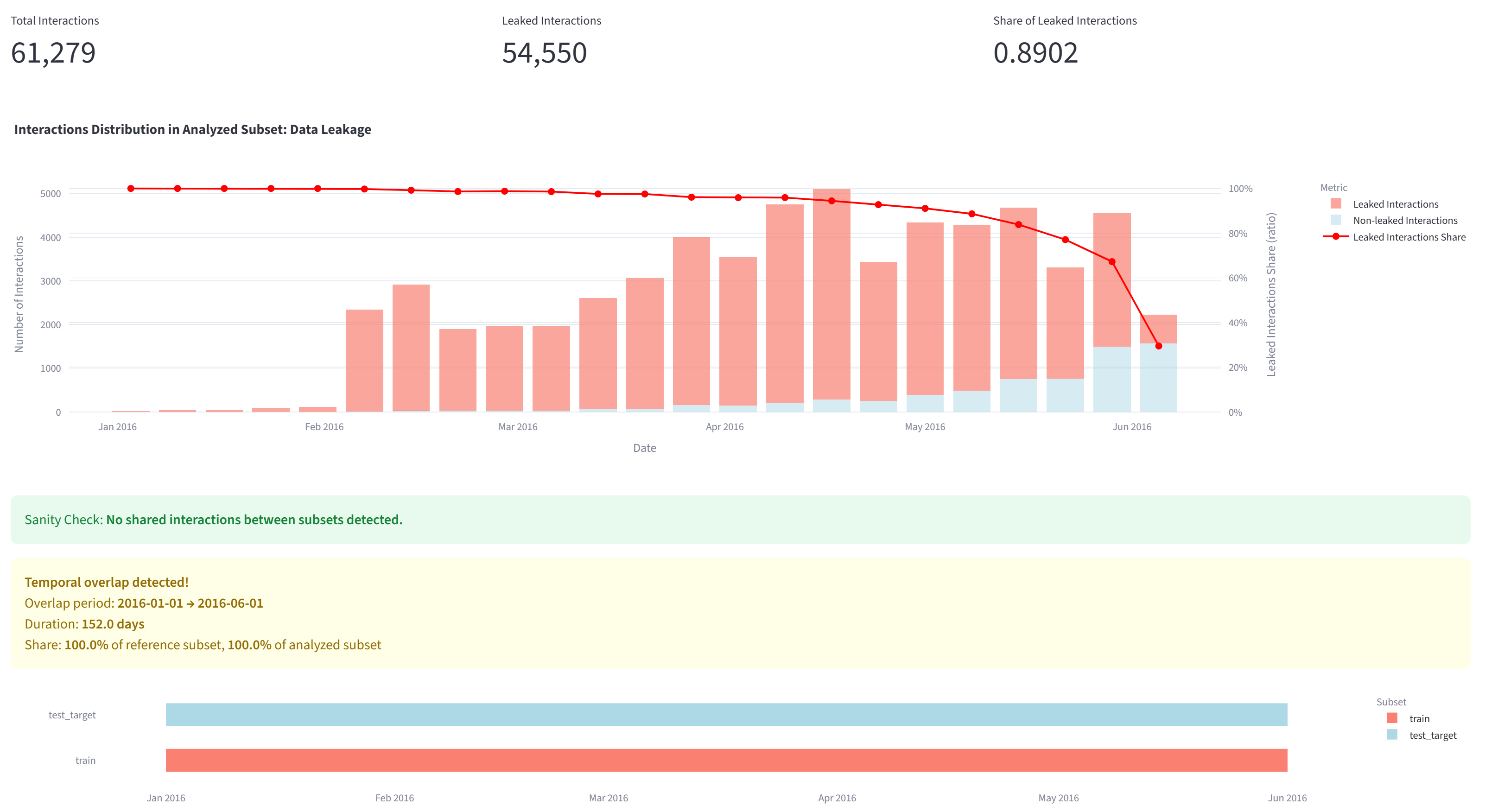}
\caption{Data leakage in Diginetica dataset (LOO): training and evaluation subsets are 100\% overlapped, 89\% targets demonstrate data leakage.} \label{fig:leakage}
\end{figure}

\begin{table}[b!]
\setlength{\abovecaptionskip}{7pt}
\caption{Core SplitLight components and visualization types}
\resizebox{1.0\columnwidth}{!}{%
\begin{tabular}{lccccccc}
\toprule
\makecell[l]{\textbf{Visualization}\\\textbf{Type}} &
\makecell{\textbf{Core \&}\\\textbf{Temporal}} &
\makecell{\textbf{Interactions}\\\textbf{Over Time}} &
\makecell{\textbf{Repeat}\\\textbf{Consumption}} &
\makecell{\textbf{Temporal}\\\textbf{Leakage}} &
\makecell{\textbf{Cold}\\\textbf{Start}} &
\makecell{\textbf{Time}\\\textbf{Deltas}} &
\makecell{\textbf{Item}\\\textbf{Positions}} \\
\midrule
Cards     & \cmark & \cmark & \cmark & \cmark & \cmark & \cmark & \cmark \\
Tabular   & \cmark &        & \cmark & \cmark & \cmark & \cmark & \cmark \\
Histogram & \cmark &        & \cmark &        &        & \cmark & \cmark \\
Timeline  &        & \cmark &        & \cmark & \cmark &        &        \\
\bottomrule
\end{tabular}
}
\label{tab:components}
\end{table}

\subsection{Summary Dashboard}

Practitioners often need a fast, repeatable sanity check before launching experiments. A compact, standardized view helps identify high-risk issues early and supports consistent reporting.

\begin{figure}[t!]

\includegraphics[width=1.0\columnwidth]{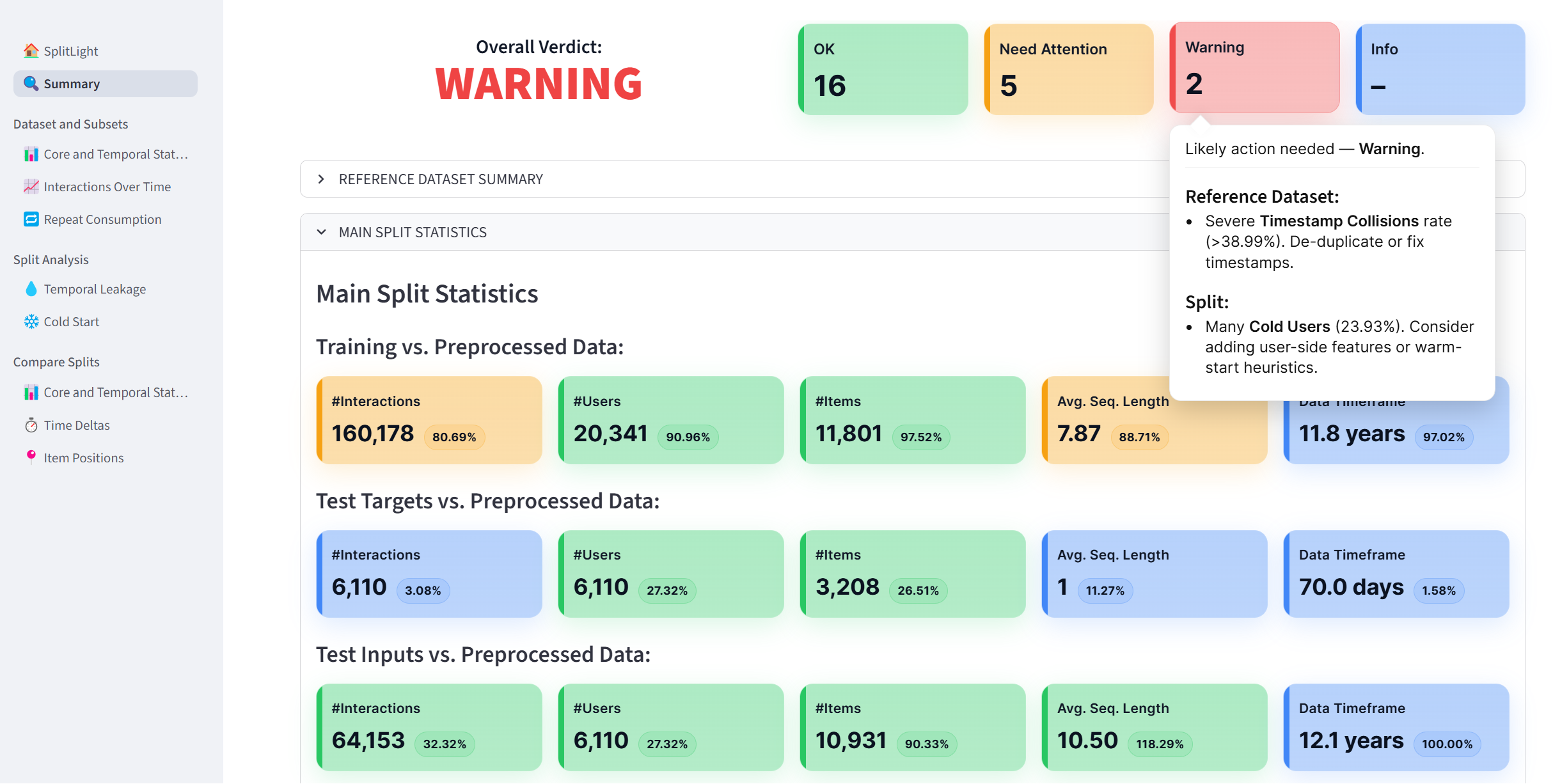}
\caption{Example of SplitLight summary dashboard showing an at-a-glance health check of the dataset and generated splits, with links to detailed diagnostics.} \label{fig:summary}
\end{figure}

\emph{Implementation.}
The Streamlit summary page consolidates the primary diagnostics into a single dashboard with configurable thresholds (Figure~\ref{fig:summary}). It is designed as an entry point: users can quickly spot potential issues and then drill down into detailed pages for the relevant component.

\subsection{Presenting Findings with SplitLight}
To summarize, SplitLight provides visualizations and reports for both high-level analysis of datasets or split artifacts and deeper analysis of specific quality aspects. The summary dashboard offers an at-a-glance overview of the dataset and split health with configurable color-coded \textit{cards} that link to corresponding detailed pages, and a data quality summary block. As shown in Table~\ref{tab:components}, the remaining pages provide statistics in \textit{tabular form}, characterize the \textit{distributions} of per-user/item/interaction aggregates using quintiles and histograms, and illustrate the data distribution \textit{timeline}. While the Streamlit interface enables standardized analysis, the code framework allows extraction of intermediate results for further processing.

\section{Case Studies}
In this section, we apply SplitLight to widespread RS datasets to identify their key properties, uncover dataset- and split-level issues, and illustrate how the identified properties may affect evaluation.

\subsection{Experimental Settings}
We use three rating datasets (MovieLens-1M/20M (ML-1M/20M), Amazon Beauty), two e-commerce clickstream datasets (Diginetica, Dressipi), and a music streaming dataset (a 20k-user subset of Zvuk). Selected datasets vary in scale, sparsity, temporal dynamics, and repeat-consumption patterns: 

For experiments described in Section~\ref{sec:res_raw}, we do not apply any preprocessing. For experiments described in Sections \ref{sec:res_split}-\ref{sec:res_model}, we \textit{preprocess} all datasets by applying 5-core filtering and consecutive duplicates removal. For users with timestamp collisions, the original event order is retained. We used the following data \textit{splits}: (i) Global temporal split at the $0.9$ quantile ($q_{0.9}$) with the all-items target strategy: the most recent 10\% of interactions form the test set, the next 10\% form the validation set, and the remaining data are used for training. We also consider (ii) leave-one-out (LOO) and (iii) GTS-$q_{0.9}$ with last-item target for experiments with RS model training. Cold items in validation and test sets are filtered by default. All splits are created after preprocessing and preserve per-user temporal order.

We train SASRec~\cite{kang2018self, tikhonovich2025esasrec} and evaluate the \textit{RS quality} using NDCG@10 (N@10), averaging results over 5 seeds. We use the above-mentioned preprocessing and data splits as a baseline setup and vary three factors: (1) keep consecutive duplicates in preprocessing (2) not preserving the within-user order of events involved in timestamp collisions, (3) not filtering cold items from evaluation subsets.
Full experimental configuration is available on Github\footnote{\url{https://github.com/monkey0head/SplitLight/tree/main/runs/configs}}.

\subsection{Raw Dataset Analysis} \label{sec:res_raw}
Tables \ref{tab:raw_dataset_stats}, \ref{tab:temp_dataset_stats} illustrate core and temporal properties of raw datasets obtained with SplitLight.
Rating datasets have longer \textit{timeframes} than industrial implicit feedback datasets (Zvuk, Dressipi, Diginetica). In turn, mean \textit{user lifetime} varies widely, from several minutes in session-based datasets (Diginetica, Dressipi) to several months in rating datasets. However, MovieLens datasets show a discrepancy: despite a long timeframe and a mean lifetime, the median user lifetime is about 1 hour, a single session, after which users rarely return. Thus, Diginetica and Dressipi reflect short-term, in-session behavior, making them well-suited for session-based model assessment, whereas MovieLens data may be non-representative for long-term preference modeling due to a skewed user lifetime distribution.

Industrial datasets are also characterized by lower mean \textit{between-interaction time} ($\Delta$t), reflecting more frequent user activity. For rating datasets, mean $\Delta$t is higher, almost reaching 70 days for the Beauty, but surprisingly, median $\Delta$t for MovieLens datasets is short and reaches 0s for ML-1M, which indicates distribution skew and possible data logging issues. One such issue is the presence of timestamp collisions, which is pronounced across all the considered rating datasets. For ML-1M, timestamp collisions occur in more than half of all interactions, making within-user temporal ordering ambiguous~\cite{hidasi2023flaws,klenitskiy2026analysis}. 
\textit{Repeat consumption} is observed across all industrial datasets, especially Zvuk (68.17\% interactions are repeats), indicating that users frequently return to their favorite tracks. The median time between interactions in raw Zvuk data is 0.25s, likely reflecting skip behavior or logging artifacts. Removing consecutive repeats (21.5\%) raises the median time between interaction $\Delta$t to 15s and reduces mean sequence length by 22.7\%, better capturing real listening patterns and more diverse user preferences. 

\begin{table}[t]
\setlength{\abovecaptionskip}{7pt}
\caption{Statistics of the raw datasets}
\label{tab:raw_dataset_stats}
\resizebox{1.0\columnwidth}{!}{%
    \centering
    \begin{tabular}{lrrrrrrrr}
    \toprule
        \textbf{Dataset} & 
        \textbf{\#Users} & 
        \textbf{\#Items} & 
        \textbf{\#Interact.} & 
        \textbf{Avg. Len.} &
        \makecell{\textbf{Density} \\ \textbf{(\%)}} & 
        \makecell{\textbf{Consec.}\\ \textbf{Repeats (\%)}} & 
        \makecell{\textbf{Repeated} \\ \textbf{Interact. (\%)}} \\
    \midrule
        Beauty~\cite{mcauley2015imagebased} & 22 363 & 12 101 & 198 502 & 8.88 & 0.07 & 0.00 & 0.00 \\
        Diginetica~\cite{diginetica2024} & 310 324 & 122 993 & 1 235 380 & 3.98 & 0.003 & 5.85 & 13.62 \\
        Dressipi~\cite{landia2022recsys} & 1 000 000 & 23 496 & 4 743 820 & 4.74 & 0.02 & 7.25 & 13.39 \\
        ML-1M~\cite{ml1} & 6 040 & 3 706 & 1 000 209 & 165.60 & 4.47 & 0.00 & 0.00 \\
        ML-20M~\cite{ml1} & 138 493 & 26 744 & 20 000 263 & 144.41 & 0.54 & 0.00 & 0.00 \\
        Zvuk~\cite{shevchenko2024variability} & 20 000 & 391 322 & 10 867 482 & 543.37 & 0.14 & 21.50 & 68.17 \\
    \bottomrule
    \end{tabular}
}
\end{table}

\subsection{Data Split Analysis} \label{sec:res_split}

For \textit{LOO split}, we observe almost 100\% train-test periods overlap and temporal data leakage for the majority of evaluation targets, as illustrated for Diginetica in Figure~\ref{fig:leakage}. We also observe the absence of cold users (as expected by split construction) and items (due to close to 100\% train-test periods overlap).

We also discovered an interaction distribution skew in ML-1M and demonstrate that \textit{GTS with a standard $q_{0.9}$} yields a target subset spanning more than 2 years (over 76\% of the dataset timeframe) (Figure \ref{fig:over_time}), which is misaligned with a real-world recommender usage scenario. We also observe that for Diginetica and Dressipi, GTS-$q_{0.9}$ yields 100\% cold users, and 82.6\% for ML-20M, making these datasets particularly suitable for evaluating cold-start models. 

\subsection{Some Effects on RS Evaluation} \label{sec:res_model}
In this section, we provide illustrative examples of how data handling can affect the offline RS evaluation outcomes (Table~\ref{tab:table_exp_res}).
We observe that for LOO split,  \textit{shuffling the interactions order for timestamp collisions} (which often happens unintentionally) \textit{leads to a consistent quality decrease up to 9.5\% across datasets}. This may indicate the presence of an artificial order of collision events. In contrast, for GTS-$q_{0.9}$ with the last-item target, we do not observe this effect, which we attribute to a less pronounced "memorization over the generalization" effect. Overall, we recommend being aware of and quantifying the collisions to determine whether affected datasets should be discarded, and whether to preserve interaction order when collisions are present.

For GTS-$q_{0.9}$ with last-item target, retaining consecutive repeats yields up to a $+60\%$ gain in NDCG@10 on datasets with a non-negligible repeat rate. This indicates that consecutive repeats are easy to memorize and strongly bias models toward short-term signals. When evaluation is restricted to non-consecutive targets, performance drops sharply, reflecting the negative effect of consecutive duplicates' presence in training data on the model's generalization.

Including cold items in evaluation subsets (both input and target) reduces NDCG@10 by $1.2$–$20\%$, correlating with the cold target share. The effect is most pronounced in datasets with many cold items (e.g., Beauty, Dressipi, and ML-20M), highlighting the need for cold start analysis and handling. 

By these experiments, we show how particular data and split properties alter model performance and metrics, making results obtained under the same split (GTS q0.9 with Last target) incompatible between papers and experiments.

\begin{table}[t!]
    \setlength{\abovecaptionskip}{7pt}
    \centering
    \caption{Temporal dataset statistics. Definitions: $\Delta t$ – time between user interactions (d = days, h = hours, m = minutes, s = seconds), Lt. – lifetime, Ts. Collis. – timestamp collisions}
    \label{tab:temp_dataset_stats}
    \setlength{\tabcolsep}{4pt}
    \footnotesize
    \resizebox{1\linewidth}{!}{%
    \small
    \begin{tabular}{lcrrrrrr}
        \toprule
        \textbf{Dataset} &
        \textbf{End Date} &
        \makecell{\textbf{Length} \\ \textbf{(years)}} &
        \makecell{\textbf{Mean} \\ $\Delta t$} &
        \makecell{\textbf{Median} \\ $\Delta t$} &
        \makecell{\textbf{Mean} \\ \textbf{User Lt.}} &
        \makecell{\textbf{Median} \\ \textbf{User Lt.}} &
        \makecell{\textbf{Ts.}  \textbf{Collis.} \\ \textbf{(\%)}}
        \\
        \midrule
        Beauty        & 2014-07-23 & 12.11 & 69.65 d & 4.00 d  & 1.50 y & 1.00 y & 38.9931 \\
        Diginetica    & 2016-06-01 & 0.42  & 1.79 m  & 1.02 m  & 5.34 m & 2.86 m & 0.0003 \\
        Dressipi      & 2021-05-31 & 1.42  & 10.02 m & 45.57 s & 37.50 m & 1.75 m & 0.0139 \\
        ML-1M         & 2003-02-28 & 2.84  & 13.85 h & 0.00 s  & 94.99 d & 0.05 d & 52.8935 \\
        ML-20M        & 2015-03-31 & 20.22 & 1.37 d  & 11.00 s & 196.59 d & 0.04 d & 20.9617 \\
        Zvuk          & 2023-04-16 & 0.25  & 1.72 h  & 0.25 s  & 38.76 d & 32.29 d & 0.2195 \\
        \bottomrule
    \end{tabular}}
\end{table}

\begin{table}[t]
  \centering
  \setlength{\abovecaptionskip}{7pt}
  \caption{Impact of different data curation protocols on RS evaluation results}
  \label{tab:table_exp_res}
  
  \renewcommand{\arraystretch}{1.15}

  \resizebox{\linewidth}{!}{%
    \begin{threeparttable}
    \begin{tabular}{
      l
      >{\large}S[table-format=1.4]
      >{\large}c
      >{\large}S[table-format=1.4]
      >{\large}S[table-format=1.4]
      >{\large}S[table-format=2.2]
      >{\large}S[table-format=1.4]
      >{\large}S[table-format=1.4]
      >{\large}S[table-format=2.2]
      >{\large}S[table-format=1.4]
    }
      \toprule

      \small 

      & \multicolumn{1}{c}{\textbf{LOO}} &
        \multicolumn{2}{c}{\textbf{Shuffle Collis., LOO}} &
        \multicolumn{1}{c}{\textbf{GTS $q_{0.9}$}} &
        \multicolumn{3}{c}{\textbf{Keep Consec. Repeats, GTS $q_{0.9}$}} &
        \multicolumn{2}{c}{\textbf{Keep Cold, GTS $q_{0.9}$}} \\
      \cmidrule(lr){2-2}
      \cmidrule(lr){3-4}
      \cmidrule(lr){5-5}
      \cmidrule(lr){6-8}
      \cmidrule(lr){9-10}

      \textbf{Dataset}
      & \multicolumn{1}{c}{\makecell{\textbf{N@10}}}
      & \multicolumn{1}{c}{\makecell{\textbf{Collis.}\\\textbf{(\%)}}}
      & \multicolumn{1}{c}{\makecell{\textbf{N@10}\\\textbf{(shuffled)}}}
      & \multicolumn{1}{c}{\makecell{\textbf{N@10}}}
      & \multicolumn{1}{c}{\makecell{\textbf{Consec.}\\\textbf{Repeats (\%)}}}
      & \multicolumn{1}{c}{\makecell{\textbf{N@10}}}
      & \multicolumn{1}{c}{\makecell{\textbf{N@10 (non- }\\\textbf{consec. targets)}}}
      & \multicolumn{1}{c}{\makecell{\textbf{Cold}\\\textbf{Targets (\%)}}}
      & \multicolumn{1}{c}{\makecell{\textbf{N@10}}} \\
      \midrule

      \Large{Beauty}     & \Large{0.0566} & \Large{38.99} & \Large{0.0521} & \Large{0.0348} & \Large{\NA}   & \Large{\NA}   & \Large{\NA}   & \Large{22.32} & \Large{0.0280} \\
      \Large{Diginetica} & \Large{0.1663} & \multicolumn{1}{c}{\normalsize $<1$} & \Large{\NA} & \Large{0.1443} & \Large{5.85} & \Large{0.2118} & \Large{0.1364} & \Large{1.50} & \Large{0.1427} \\
      \Large{Dressipi}   & \Large{0.1318} & \multicolumn{1}{c}{\normalsize $<1$} & \Large{\NA} & \Large{0.1343} & \Large{7.25} & \Large{0.2168} & \Large{0.1225} & \Large{32.59} & \Large{0.0965} \\
      \Large{ML-1M}      & \Large{0.1861} & \Large{52.89} & \Large{0.1685} & \Large{0.0905} & \Large{\NA}   & \Large{\NA}   & \Large{\NA}   & \Large{0.50} & \Large{0.0907} \\
      \Large{ML-20M}     & \Large{0.1601} & \Large{20.96} & \Large{0.1469} & \Large{0.0836} & \Large{\NA}   & \Large{\NA}   & \Large{\NA}   & \Large{28.40} & \Large{0.0704} \\
      \Large{Zvuk}       & \Large{0.2594} & \multicolumn{1}{c}{\normalsize $<1$} & \Large{\NA} & \Large{0.1529} & \Large{21.50} & \Large{0.2653} & \Large{0.1247} & \Large{4.61} & \Large{0.1362} \\

      \bottomrule
    \end{tabular}
    \end{threeparttable}
}
\end{table}

\section{Conclusion}

In this work, we summarized essential dataset properties and the data curation choices derived from previous research and practical \textit{lessons learned}. We highlight the need to (1) verify dataset and split validity and closeness to intended RS usage, (2) quantify and transparently handle data issues, repeats and timestamp anomalies, (3) measure and report cold-start exposure in line with the intended deployment regime, (4) check absence of data leakage from future, and (5) ensure evaluation subsets adequately represent the overall distribution. Across six datasets, we showed that minor curation decisions can materially change evaluation outcomes. 
We introduced \textit{SplitLight}, an open-source toolkit that makes data properties, and preprocessing and splitting choices \textit{visible and testable} via a Python toolkit and an interactive UI with audit summaries and split comparisons. SplitLight is available for the community under the MIT License and ready to be applied to make dataset and splits analysis easier, improve reproducibility, and cross-paper comparability.

\bibliographystyle{ACM-Reference-Format}
\bibliography{bibliography}

@inproceedings{mancino2025datarec,
  title={DataRec: A Python library for standardized and reproducible data management in recommender systems},
  author={Mancino, Alberto Carlo Maria and Bufi, Salvatore and Di Fazio, Angela and Ferrara, Antonio and Malitesta, Daniele and Pomo, Claudio and Di Noia, Tommaso},
  booktitle={Proceedings of the 48th International ACM SIGIR Conference on Research and Development in Information Retrieval},
  pages={3478--3487},
  year={2025}
}

@inproceedings{shevchenko2024variability,
  title={From variability to stability: Advancing RecSys benchmarking practices},
  author={Shevchenko, Valeriy and Belousov, Nikita and Vasilev, Alexey and Zholobov, Vladimir and Sosedka, Artyom and Semenova, Natalia and Volodkevich, Anna and Savchenko, Andrey and Zaytsev, Alexey},
  booktitle={Proceedings of the 30th ACM SIGKDD Conference on Knowledge Discovery and Data Mining},
  pages={5701--5712},
  year={2024}
}

@inproceedings{klenitskiy2024does,
  title={Does it look sequential? an analysis of datasets for evaluation of sequential recommendations},
  author={Klenitskiy, Anton and Volodkevich, Anna and Pembek, Anton and Vasilev, Alexey},
  booktitle={Proceedings of the 18th ACM Conference on Recommender Systems},
  pages={1067--1072},
  year={2024}
}

@inproceedings{gusak2025time,
  title={Time to Split: Exploring Data Splitting Strategies for Offline Evaluation of Sequential Recommenders},
  author={Gusak, Danil and Volodkevich, Anna and Klenitskiy, Anton and Vasilev, Alexey and Frolov, Evgeny},
  booktitle={Proceedings of the Nineteenth ACM Conference on Recommender Systems},
  pages={874--883},
  year={2025}
}

@article{ml1,
  title={The movielens datasets: History and context},
  author={Harper, F Maxwell and Konstan, Joseph A},
  journal={Acm transactions on interactive intelligent systems (tiis)},
  volume={5},
  number={4},
  pages={1--19},
  year={2015},
  publisher={Acm New York, NY, USA}
}

@misc{mcauley2015imagebased,
  title={Image-based recommendations on styles and substitutes},
  author={McAuley, Julian and Targett, Christopher and Shi, Qinfeng and Van Den Hengel, Anton},
  booktitle={Proceedings of the 38th international ACM SIGIR conference on research and development in information retrieval},
  pages={43--52},
  year={2015}
}

@inproceedings{cold2024,
author = {Chen, Gaode and Sun, Ruina and Jiang, Yuezihan and Cao, Jiangxia and Zhang, Qi and Lin, Jingjian and Li, Han and Gai, Kun and Zhang, Xinghua},
title = {A Multi-modal Modeling Framework for Cold-start Short-video Recommendation},
year = {2024},
isbn = {9798400705052},
publisher = {Association for Computing Machinery},
address = {New York, NY, USA},
doi = {10.1145/3640457.3688098},
booktitle = {Proceedings of the 18th ACM Conference on Recommender Systems},
pages = {391–400},
series = {RecSys '24}
}

@article{sukhorukov2025maximum,
  title={Maximum Impact with Fewer Features: Efficient Feature Selection for Cold-Start Recommenders through Collaborative Importance Weighting},
  author={Sukhorukov, Nikita and Gusak, Danil and Frolov, Evgeny},
  journal={arXiv preprint arXiv:2508.06455},
  year={2025}
}

@inproceedings{hidasi2023flaws,
  title = {Widespread Flaws in Offline Evaluation of Recommender Systems},
  author = {Hidasi, Bal{\'a}zs and Czapp, {\'A}d{\'a}m Tibor},
  booktitle = {Proceedings of the 17th ACM Conference on Recommender Systems (RecSys ’23)},
  year = {2023},
  url = {https://hidasi.eu/publications/}
}

@inproceedings{mezentsev2024scalable,
  title={Scalable cross-entropy loss for sequential recommendations with large item catalogs},
  author={Mezentsev, Gleb and Gusak, Danil and Oseledets, Ivan and Frolov, Evgeny},
  booktitle={Proceedings of the 18th ACM Conference on Recommender Systems},
  pages={475--485},
  year={2024}
}

@inproceedings{meng2020splits,
  title={Exploring data splitting strategies for the evaluation of recommendation models},
  author={Meng, Zaiqiao and McCreadie, Richard and Macdonald, Craig and Ounis, Iadh},
  booktitle={Proceedings of the 14th acm conference on recommender systems},
  pages={681--686},
  year={2020}
}

@article{volodkevich2025autoregressive,
  title={Autoregressive generation strategies for Top-K sequential recommendations},
  author={Volodkevich, Anna and Gusak, Danil and Klenitskiy, Anton and Pembek, Anton and Vasilev, Alexey},
  journal={User Modeling and User-Adapted Interaction},
  volume={35},
  number={3},
  pages={13},
  year={2025},
  publisher={Springer}
}

@article{ji2020leakage,
  title = {A Critical Study on Data Leakage in Recommender System Offline Evaluation},
  author={Ji, Yitong and Sun, Aixin and Zhang, Jie and Li, Chenliang},
  journal={ACM Transactions on Information Systems},
  volume={41},
  number={3},
  pages={1--27},
  year={2023},
  publisher={ACM New York, NY}
}

@article{ludewig2018sessioneval,
  title = {Evaluation of Session-based Recommendation Algorithms},
  author = {Ludewig, Malte and Jannach, Dietmar},
  journal = {User Modeling and User-Adapted Interaction},
  year = {2018},
  doi = {10.1007/s11257-018-9209-6},
}

@inproceedings{sun2023take,
  title={Take a fresh look at recommender systems from an evaluation standpoint},
  author={Sun, Aixin},
  booktitle={Proceedings of the 46th International ACM SIGIR Conference on Research and Development in Information Retrieval},
  pages={2629--2638},
  year={2023}
}

@inproceedings{le2025don,
  title={Don't Get Ahead of Yourself: A Critical Study on Data Leakage in Offline Evaluation of Sequential Recommenders},
  author={Le, Huy Hoang and Liu, Yang and Medlar, Alan and Glowacka, Dorota},
  booktitle={Proceedings of the Nineteenth ACM Conference on Recommender Systems},
  pages={1164--1168},
  year={2025}
}

@article{fan2024our,
  title={Our model achieves excellent performance on MovieLens: what does it mean?},
  author={Fan, Yu-chen and Ji, Yitong and Zhang, Jie and Sun, Aixin},
  journal={ACM Transactions on Information Systems},
  volume={42},
  number={6},
  pages={1--25},
  year={2024},
  publisher={ACM New York, NY, USA}
}

@inproceedings{tan2016improved,
  title     = {Improved Recurrent Neural Networks for Session-based Recommendations},
  author    = {Tan, Yong Kiam and Xu, Xinxing and Liu, Yong},
  booktitle = {Proceedings of the 1st Workshop on Deep Learning for Recommender Systems (DLRS@RecSys '16)},
  year      = {2016},
  pages     = {17--22},
  doi       = {10.1145/2988450.2988452}
}

@inproceedings{xie2022cl4srec,
  title     = {Contrastive Learning for Sequential Recommendation},
  author    = {Xie, Xu and Sun, Fei and Liu, Zhaoyang and Wu, Shiwen and Gao, Jinyang and Zhang, Jiandong and Ding, Bolin and Cui, Bin},
  booktitle = {Proceedings of the IEEE 38th International Conference on Data Engineering (ICDE)},
  year      = {2022},
  pages     = {1259--1273},
  doi       = {10.1109/ICDE53745.2022.00099}
}

@inproceedings{sachdeva2022samplingcf,
  title     = {On Sampling Collaborative Filtering Datasets},
  author    = {Sachdeva, Noveen and Wu, Carole-Jean and McAuley, Julian},
  booktitle = {Proceedings of the Fifteenth ACM International Conference on Web Search and Data Mining (WSDM '22)},
  year      = {2022},
  doi       = {10.1145/3488560.3498439}
}

@inproceedings{gusak2025dish,
  title={Recommendation Is a Dish Better Served Warm},
  author={Gusak, Danil and Sukhorukov, Nikita and Frolov, Evgeny},
  booktitle={Proceedings of the Nineteenth ACM Conference on Recommender Systems},
  doi={10.1145/3705328.3759331},
  year={2025}
}

@inproceedings{pembek2025let,
  title={Let It Go? Not Quite: Addressing Item Cold Start in Sequential Recommendations with Content-Based Initialization},
  author={Pembek, Anton and Fatkulin, Artem and Klenitskiy, Anton and Vasilev, Alexey},
  booktitle={Proceedings of the Nineteenth ACM Conference on Recommender Systems},
  pages={626--631},
  year={2025}
}

@inproceedings{repeatconsumption2020doi,
  title = {Explainable Recommendation for Repeat Consumption},
  year  = {2020},
  author = {Tsukuda, Kosetsu and Goto, Masataka},
  doi   = {10.1145/3383313.3412230}
}

@article{massey1951ks,
  author  = {Frank J. Massey, Jr.},
  title   = {The Kolmogorov-Smirnov Test for Goodness of Fit},
  journal = {Journal of the American Statistical Association},
  year    = {1951},
  volume  = {46},
  number  = {253},
  pages   = {68--78}
}

@article{klenitskiy2026analysis,
  title={An Analysis of Sequential Patterns in Datasets for Evaluation of Sequential Recommendations},
  author={Klenitskiy, Anton and Volodkevich, Anna and Pembek, Anton and Vasilev, Alexey},
  journal={ACM Transactions on Recommender Systems},
  year={2026},
  publisher={ACM New York, NY}
}

@inproceedings{better2024,
  title={Better generalization with semantic ids: A case study in ranking for recommendations},
  author={Singh, Anima and Vu, Trung and Mehta, Nikhil and Keshavan, Raghunandan and Sathiamoorthy, Maheswaran and Zheng, Yilin and Hong, Lichan and Heldt, Lukasz and Wei, Li and Tandon, Devansh and others},
  booktitle={Proceedings of the 18th ACM Conference on Recommender Systems},
  pages={1039--1044},
  year={2024}
}

@incollection{landia2022recsys,
  title={Recsys challenge 2022 dataset: Dressipi 1m fashion sessions},
  author={Landia, Nick and Mcalister, Rachael and North, Donna and Kalloori, Saikishore and Srivastava, Abhishek and Ferwerda, Bruce},
  booktitle={Proceedings of the Recommender Systems Challenge 2022},
  pages={1--3},
  year={2022}
}

@inproceedings{kang2018self,
  title={Self-attentive sequential recommendation},
  author={Kang, Wang-Cheng and McAuley, Julian},
  booktitle={2018 IEEE international conference on data mining (ICDM)},
  pages={197--206},
  year={2018},
  organization={IEEE}
}

@dataset{diginetica2024,
  author       = {Garg, Diksha and Gupta, Priyanka and Malhotra, Pankaj and Vig, Lovekesh and Shroff, Gautam},
  title        = {Diginetica},
  year         = {2024},
  doi          = {10.57702/t18y8j18},
  url          = {https://doi.org/10.57702/t18y8j18}
}

@inproceedings{tikhonovich2025esasrec,
  title={eSASRec: Enhancing Transformer-based Recommendations in a Modular Fashion},
  author={Tikhonovich, Daria and Zelinskiy, Nikita and Petrov, Aleksandr V and Spirina, Mayya and Semenov, Andrei and Savchenko, Andrey V and Kuliev, Sergei},
  booktitle={Proceedings of the Nineteenth ACM Conference on Recommender Systems},
  pages={1175--1180},
  year={2025}
}

@article{khrylchenko2025scaling,
  title={Scaling recommender transformers to one billion parameters},
  author={Khrylchenko, Kirill and Matveev, Artem and Makeev, Sergei and Baikalov, Vladimir},
  journal={arXiv preprint arXiv:2507.15994},
  year={2025}
}

\end{document}